\documentclass[aps,pra,twocolumn,groupedaddress,nofootinbib,showpacs]{revtex4-1}
\usepackage{amsmath}
\usepackage[pdftex]{hyperref}
\usepackage{graphicx}
\usepackage{float}
\usepackage[abs]{overpic}
\usepackage{color}
\usepackage{subfigure}
\begin{document}

\title{Variational study of polarons and bipolarons in a 1D Bose lattice gas in both superfluid and Mott regimes}

\author{Shovan Dutta}
\email{sd632@cornell.edu}
\author{Erich J. Mueller}
\email{em256@cornell.edu}
\affiliation{Laboratory of Atomic and Solid State Physics, Cornell University, Ithaca, New York 14850, USA}

\date{\today}

\begin{abstract}
We use variational methods to study a spin impurity in a 1D Bose lattice gas. Both in the strongly interacting superfluid regime and the Mott regime we find that the impurity binds with a hole, forming a polaron. Our calculations for the dispersion of the polaron are consistent with recent experiments by Fukuhara {\it et. al.} [\href{http://www.nature.com/nphys/journal/v9/n4/full/nphys2561.html}{Nature Phys. {\bf 9}, 235 (2013)}] and give a better understanding of their numerical simulations. We find that for sufficiently weak interactions there are ranges of momentum for which the polaron is unstable. We propose experimentally studying the stability of the polaron by measuring the correlation between the impurity and holes. We also study two interacting impurities, finding stable bipolarons for sufficiently strong interactions.
\end{abstract}

\pacs{67.85.Hj, 67.85.De, 67.30.hj, 71.38.-k}

\maketitle

\section{Introduction}\label{intro}
\vspace{-0.02cm}
Using single-site imaging techniques \cite{greiner 2002,greiner 2009,greiner 2010,shersen,weitenberg,gadway} it is now possible to track the motion of spin impurities in a gas of cold atoms trapped in an optical lattice \cite{palzer,schirotzek,catani,fukuhara,fukuhara2}. Such a direct probe is unprecedented in condensed matter physics \cite{bloch_nature,lewenstein,bloch,focus}, and has stimulated a rich body of theoretical work \cite{astrakharchik,gangardt,cazalilla,johnson,yariv,massel}. These experimental and theoretical studies are motivated in part by connections between the physics of a spin impurity and larger questions in quantum magnetism \cite{greiner 2011}, high-$\mbox{T}_{\mbox{{\scriptsize{c}}}}$ superconductivity \cite{klein}, and transition-metal oxides \cite{imada}. Here we present a theoretical study of the properties of spin impurities in a 1D Bose lattice gas.

In a typical experiment of this type, one first prepares an array of spin-polarized atoms on a lattice. Then Raman lasers flip one or more of these atomic spins, creating spin impurities. The excitations of the spin-polarized bath can dress such an impurity, producing a composite particle called a polaron \cite{alexandrov,feynman,steffen}. In one recent experimental study involving a bosonic spin impurity in ${}^{87}$Rb, Fukuhara {\it et. al.} found preliminary evidence of polaron-like behavior within the superfluid regime \cite{fukuhara}. They observed a suppression of the bath density near the impurity, and a strong renormalization of the impurity's hopping rate. In the Mott phase, their results are understood by mapping the system to a Heisenberg chain \cite{garcia-ripoll,giamarchi,kuklov,demler}, whereas in the superfluid phase, they find good agreement with numerical t-DMRG simulations \cite{tdmrg}. Here we use simple variational arguments to explain the underlying physics.

We model this system by the two-species Bose-Hubbard Hamiltonian \cite{sachdev,jaksch}. In Sec. \ref{physical system}, we analytically study the limiting cases of very strong and very weak coupling. Guided by these limiting behaviors, in Sec. \ref{variational wavefunction subsection}, we propose a simple variational model that captures the physics in both limits, extending those descriptions to all interaction strengths. Our model begins with the Gutzwiller mean-field wavefunction \cite{roshkar,krauth,fisher,sheshadri}, and adds correlation between a single impurity and a hole. We find that our ansatz provides a rich picture of the physics of a spin impurity, and we believe it fully captures all the relevant physics. It is exact in the strongly and weakly interacting limits, but, like the underlying Gutzwiller mean-field theory, we do not believe that it is quantitatively accurate for intermediate coupling \cite{zwerger,zoller}.

In terms of the single particle hopping rate $J$ and the on-site interaction $U$ (see Eq. (\ref{bose-hubbard})), we find stable polarons for all momenta when $J/U \lesssim 2.3$. This agrees with the experimental observation of a stable polaron at $J/U = 0.47$ \cite{fukuhara}. We fully characterize the polaron, calculating its energy, spatial structure, and dispersion. From the dispersion we calculate the rate of expansion for a wave-packet, and find qualitative agreement with experimental and numerical studies in Ref. \cite{fukuhara}. At weaker coupling ($J/U \gtrsim 2.3$) our ansatz predicts that the energy for a total momentum $k$ may be lowered by unbinding the hole from the impurity. For $J/U \approx 2.3$, this instability only occurs for $k \approx 2\pi/3a$, where $a$ is the lattice spacing. As $J/U$ is increased, the instability window grows. Future experiments can map out such a ``polaron phase diagram" by studying the correlations between the impurity and the density of the bath. We provide detailed predictions for such measurements.

Adding a second impurity to the system admits the possibility of a bound state of two polarons, a bipolaron. Such bound states are of intrinsic interest for a variety of reasons, including their possible role in high-$\mbox{T}_{\mbox{{\scriptsize{c}}}}$ superconductivity \cite{fukuhara2,multipolaron,mott}. In a recent experiment, two-magnon bound states were observed in the Mott phase \cite{fukuhara2}. The measurements are consistent with analytical predictions of the Heisenberg model. The study of polaron binding in the superfluid phase is much more challenging, and has not previously been explored in detail. We study a simple generalization of our original variational model for the case of two impurities with zero total momentum. Our results indicate the formation of stable bipolarons in the superfluid phase for sufficiently strong interactions.

The rest of this article is organized as follows. In Sec. \ref{formalism}, we introduce the physical system and describe our proposed variational model. We analyze the system's properties in the Mott and the deep superfluid regime, with emphasis on how the correlation length of the impurity-hole binding changes with interaction strength. In Sec. \ref{results}, we discuss several physical predictions of our model, and present numerical results. In particular, we identify two qualitatively distinct regions in the superfluid phase, polaronic and ``two-particle." We show how the crossover can be detected experimentally from correlation measurements. Our variational model is extended to incorporate two impurities in Sec. \ref{bipolaron section} where we infer the existence of stable bipolarons at adequately large interactions. Finally, we summarize our findings and indicate possible directions of future research in Sec. \ref{conclusions}. The appendices contain derivations of key analytical results.
\section{Formalism}\label{formalism}
\subsection{The physical system and its limiting behaviors}\label{physical system}
We consider a one-dimensional chain of bosonic atoms in an optical lattice with a single spin impurity. Such a system can be experimentally realized by initially preparing the atoms (e.g., ${}^{87}$Rb) in a definite hyperfine state (such as $\vert F=1,m_F=-1\rangle$), and then changing the hyperfine state of one atom by single-site addressing technique (for example, to $\vert F=2,m_F=-2\rangle$) \cite{fukuhara}. The system is described by the two-species single-band Bose-Hubbard Hamiltonian at unity filling \cite{sachdev,jaksch}:
\begin{equation}
\hat{H} = -J \sum_{(l_1,l_2),\sigma} \hat{b}_{l_1,\sigma}^{\dagger} \hat{b}_{l_2,\sigma} + \frac{U}{2} \sum_{l,\sigma,\sigma'} \hat{n}_{l,\sigma} \hat{n}_{l,\sigma'} - \mu \sum_{l,\sigma} \;\hat{n}_{l,\sigma}\;.
\label{bose-hubbard}
\end{equation}
Here $(l_1,l_2)$ varies over all neighboring sites $l_1$ and $l_2$, $\sigma$ denotes the spin-index (`$\uparrow$' or `$\downarrow$'), $J$ represents the single-particle hopping amplitude, and $U$ is the on-site repulsion energy. As is appropriate for models of ${}^{87}$Rb, the interactions only depend on the total density on a site, and not the density of each spin component. $\hat{b}_{l,\sigma}^{\dagger}$ ($\hat{b}_{l,\sigma}$) and $\hat{n}_{l,\sigma}$ denote the creation (annihilation) and number operators for the boson of spin $\sigma$ at site $l$. The chemical potential $\mu$ should be chosen so that the ground state is at unity filling. Although the experiment includes an additional trap along the chain, we do not model it here, as all observations are made near the center of the trap where the potential is roughly constant. The system undergoes a Mott-superfluid phase transition as $J/U$ is increased beyond a critical value, $(J/U)_c$ $\approx 0.086$ within mean-field theory \cite{sachdev}. In comparing with experiments it is useful to note that the Gutzwiller ansatz overestimates the stability of the superfluid, and the Mott transition actually occurs at $J/U \approx 0.29$ \cite{zakrzewski}.

\paragraph{{\bf Mott regime}} For $J \ll U$, single-particle hopping is energetically expensive, as it changes the on-site populations. This results in an interaction driven ``Mott" insulator. However, the impurity is able to move through a second order process, and the system can be mapped onto the isotropic spin-1/2 Heisenberg chain \cite{garcia-ripoll,giamarchi,kuklov,demler}
\begin{equation}
\hat{H}_{\mbox{\scriptsize{eff}}} = -\frac{J_{\mbox{\scriptsize{ex}}}}{2} \sum_{(i,j)} (\hat{S}_i^+ \hat{S}_j^- + \hat{S}_i^- \hat{S}_j^+) - 
J_{\mbox{\scriptsize{ex}}} \sum_{(i,j)} \hat{S}_i^z \hat{S}_j^z\;,
\label{heisenberg}
\vspace{-0.1cm}
\end{equation}
where $\hat{S}_i^+ = \vert\uparrow\rangle_i \vert\downarrow\rangle_i$ and $\hat{S}_i^- = \vert\downarrow\rangle_i \vert\uparrow\rangle_i$ are the spin-flip operators, $\hat{S}_i^z = (\hat{n}_{i,\uparrow} - \hat{n}_{i,\downarrow})/2$, and $J_{\mbox{\scriptsize{ex}}} = 4 J^2 / U$ is the superexchange coupling. Here the impurity has dispersion
\begin{equation}
\varepsilon_{\mbox{\scriptsize{Mott}}}(k) = \varepsilon_{\mbox{\scriptsize{Mott}}}(0) + J_{\mbox{\scriptsize{ex}}} (1 - \cos{k})
\label{effective dispersion}
\end{equation}
corresponding to eigenstates
\begin{equation}
|k_{\mbox{\scriptsize{Mott}}}\rangle \hspace{-0.05cm}=\hspace{-0.1cm} \sum_j \hspace{-0.05cm} e^{i k j} \Big[\vert\downarrow\rangle_j + \frac{J}{U} \big\{(1 + e^{i k}) \vert + \rangle_j + (1 + e^{-i k}) \vert - \rangle_j \big\}\Big],
\label{Mott state}
\end{equation}
where $\vert\downarrow\rangle_j$ is the state where the `$\downarrow$' impurity is localized at site $j$, and $\vert \pm \rangle_j = \hat{b}_{j\pm1,\uparrow} \hat{b}_{j,\uparrow}^{\dagger} \vert\downarrow\rangle_j$ (see Appendix \ref{appendix a} for a derivation). We see from Eq. (\ref{Mott state}) that the correlation-hole is mostly localized at the impurity site, with a spread of order $(J/U)^2$ into the neighboring sites.

\paragraph{{\bf Deep superfluid regime}} In the weak coupling limit ($U \ll J$), one can study the system within the Bogoliubov approximation \cite{bogoliubov,mahan}, where one takes quadratic fluctuations about a state where $\hat{b}_{0,\sigma} = \hat{b}_{0,\sigma}^{\dagger} = \sqrt{N^{\sigma}}$, $N^{\sigma}$ being the number of particles in the condensate of spin $\sigma$. The single impurity physics emerges in the limit $N^{\downarrow} \to 1$.

The Bose-Hubbard Hamiltonian (Eq. (\ref{bose-hubbard})) can be expressed in momentum-space as ($\mathcal{N}$ denotes the total number of lattice sites)
\begin{align}
\nonumber \hat{H} =& - \sum_{p,\sigma} \Big(2 J \cos{p} + \mu - \frac{U}{2}\Big)\; \hat{b}_{p,\sigma}^{\dagger} \hat{b}_{p,\sigma} \\
&+ \frac{U}{2 \mathcal{N}} \sum_{p_1,p_2,q,\sigma_1,\sigma_2} \hat{b}_{p_1,\sigma_1}^{\dagger} \hat{b}_{p2,\sigma_2}^{\dagger} \hat{b}_{p1+q,\sigma_1} \hat{b}_{p_2-q,\sigma_2}\;,
\label{momentum space hamiltonian}
\end{align}
where the momenta are summed over $2\pi m/\mathcal{N}$ with integer $m$. To quadratic order in fluctuations, (see Appendix \ref{appendix b} for derivation)
\begin{align}
\hat{H} &= E_0 + \sum_{p \neq 0} (\varepsilon_c (p) \;\hat{c}_p^{\dagger} \hat{c}_p + \varepsilon_0 (p) \;\hat{d}_p^{\dagger} \hat{d}_p)\;,
\label{bogoliubov hamiltonian} \\
\mbox{where} \quad &\varepsilon_0 (p) = 2 J \;(1-\cos{p}) \;, \label{epsilon0}\\
&\varepsilon_c (p) = \sqrt{(\varepsilon_0 (p))^2 + 2 \;\varepsilon_0 (p) \;U\; (n^{\uparrow} + n^{\downarrow})} \label{epsilonc}
\end{align}
are the excitation spectra, $n^{\sigma}$ denote the average particle densities of the condensates, and $E_0$ is a constant. $\hat{c}_p$ and $\hat{d}_p$ are the annihilation operators of the Bogoliubov quasi-particles, defined by the canonical transformation
\begin{align}
&\hat{b}_{p,\uparrow/\downarrow} = \sqrt{\frac{n^{\uparrow/\downarrow}}{n^{\uparrow} + n^{\downarrow}}} (u_p \hat{c}_p + v_p \hat{c}_{-p}^{\dagger}) \mp \sqrt{\frac{n^{\downarrow/\uparrow}}{n^{\uparrow} + n^{\downarrow}}} \;\hat{d}_p\;,
\label{bogoliubov transformation} \\
&\mbox{with} \; u_p,v_p = 0.5 \big[\sqrt{\varepsilon_0 (p) / \varepsilon_c (p)} \pm \sqrt{\varepsilon_c (p) / \varepsilon_0 (p)}\;\big].
\label{uk and vk}
\end{align}

\begin{figure}[h]
\raggedleft
\begin{overpic}[trim=0cm 0cm 0.5cm 0cm, clip=true, width=82\unitlength]{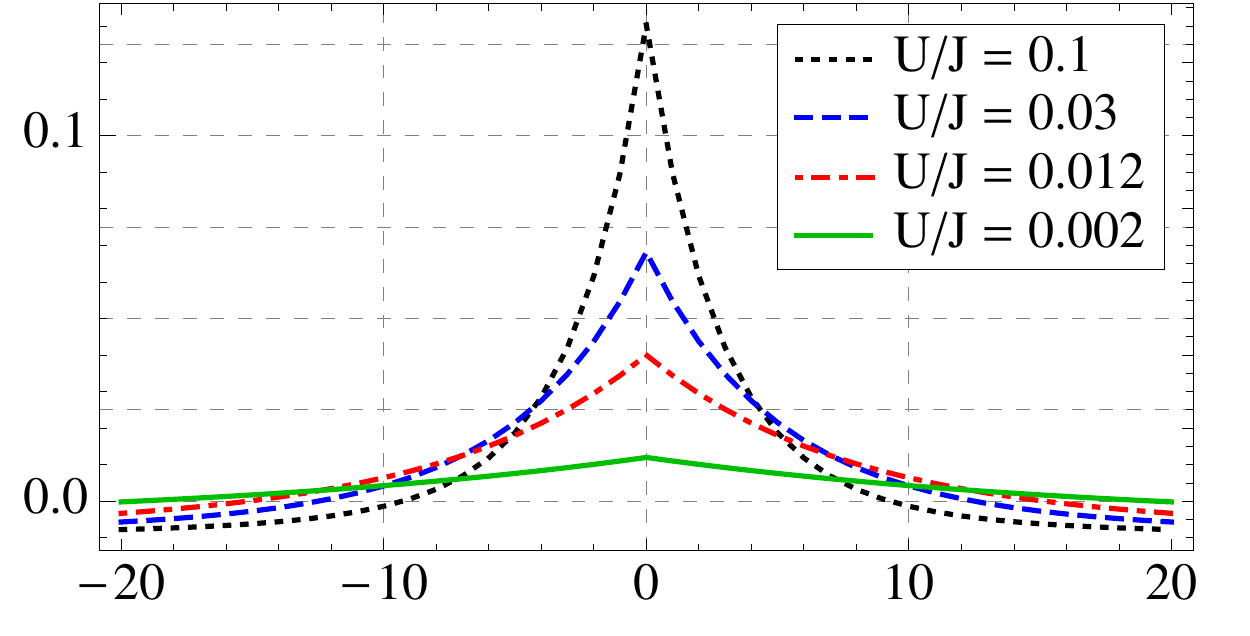}
  \put(-3,7.5){\rotatebox{90}{Correlation-hole density}}
  \put(36,-1.7){lattice sites}
\end{overpic}
\vspace{1\unitlength}
\caption{(Color online) Correlation-hole density in the Bogoliubov approximation for zero total momentum, plotted with $\mathcal{N}=101$. The impurity-hole binding weakens at lower $U/J$, leading to a flatter profile.}
\label{hole density bogoliubov}
\end{figure}

In the limit $n^{\downarrow} \to 0$ and $n^{\uparrow} \to 1$, we wish to calculate $n_{h,j}$, the density of holes at a distance $j$ from the impurity. We relate $n_{h,j}$ to a correlation function by noting that in the limit of small $n^{\downarrow}$,
\begin{equation}
C_j \equiv \langle \hat{b}_{j,\uparrow}^{\dagger} \hat{b}_{j,\uparrow} \hat{b}_{0,\downarrow}^{\dagger} \hat{b}_{0,\downarrow}\rangle = n^{\downarrow} (1 - n_{h,j})\;.
\label{cj}
\end{equation}
Direct calculation of $C_j$ then yields
\begin{equation}
n_{h,j} = \frac{1}{\mathcal{N}} \sum_{p \neq 0} \bigg(1-\frac{\varepsilon_0 (p)}{\varepsilon_c (p)}\bigg) \cos{p j}\;.
\label{bogoliubov correlation}
\end{equation}
As shown in Fig. \ref{hole density bogoliubov}, there is a strong tendency to have a hole near the impurity.

As indicated by the strength of the correlations at short distances, the impurity-hole binding becomes weaker at lower interaction. The area under the curves in Fig. \ref{hole density bogoliubov} are constant. In fact, summing Eq. (\ref{bogoliubov correlation}) over all $j$ yields $\sum_j n_{h,j}=0$. The impurity ``pushes away" the bath atoms, causing an excess of particles far away.
\subsection{The variational wavefunction}\label{variational wavefunction subsection}
Guided by the limiting properties of the impurity-hole binding discussed above, we propose the following variational wavefunction for the system with momentum $k$:
\begin{align}
\nonumber \vert k \rangle &= \sum_j \vert j \rangle \;e^{i k j}\;,\;\;\mbox{where}\\
\vert j \rangle &= \sum_i \Big(f_i \;\hat{b}_{i+j,\uparrow} \hat{b}_{j,\downarrow}^{\dagger} \vert MF \rangle\Big) + A \;\hat{b}_{j,\downarrow}^{\dagger} \vert MF \rangle\;.
\label{variational wavefunction}
\end{align}
Here $\vert MF \rangle = \prod_l \sum_n \beta_n \vert n \rangle_l$ denotes the Gutzwiller mean-field ground state of the bath, where the amplitudes $\beta_n$ for having $n$ bath atoms on a site are determined by minimizing the energy. Variational parameters $A$ and $f_i$ encode whether and how strongly the impurity binds with holes at different distances.

In the Mott phase, the impurity is strongly bound to a localized hole with a small spread, as seen from Eq. (\ref{Mott state}). Thus in this limit we expect $A \to 0$, $f_{\pm 1} \to (J/U)(1 + e^{\pm i k}) f_0$, and $f_i \lesssim \mathcal{O}((J/U)^2)$ for $|i| \geq 2$. Whereas for weak coupling, the $f_i$ s should approach uniform magnitudes as the interactions are lowered, since the correlation length ought to increase. These conjectures are confirmed in our numerical studies. In the next section, we present several physical predictions of our model. For our numerical calculation we use 101 lattice sites with a maximum of 20 bath atoms at one site. Throughout the remainder we set $\hbar=1$ and $a=1$. We label the optimized energy of the variational state as $E_{\mbox{\scriptsize{var}}}(J/U,k)$.
\section{Results}\label{results}
\subsection{Polarons}\label{polaron section}
\begin{figure}[h]
\raggedleft
\begin{overpic}[trim=0cm 0cm 1cm 0cm, clip=true, width=80\unitlength, height= 38\unitlength ,tics=2]{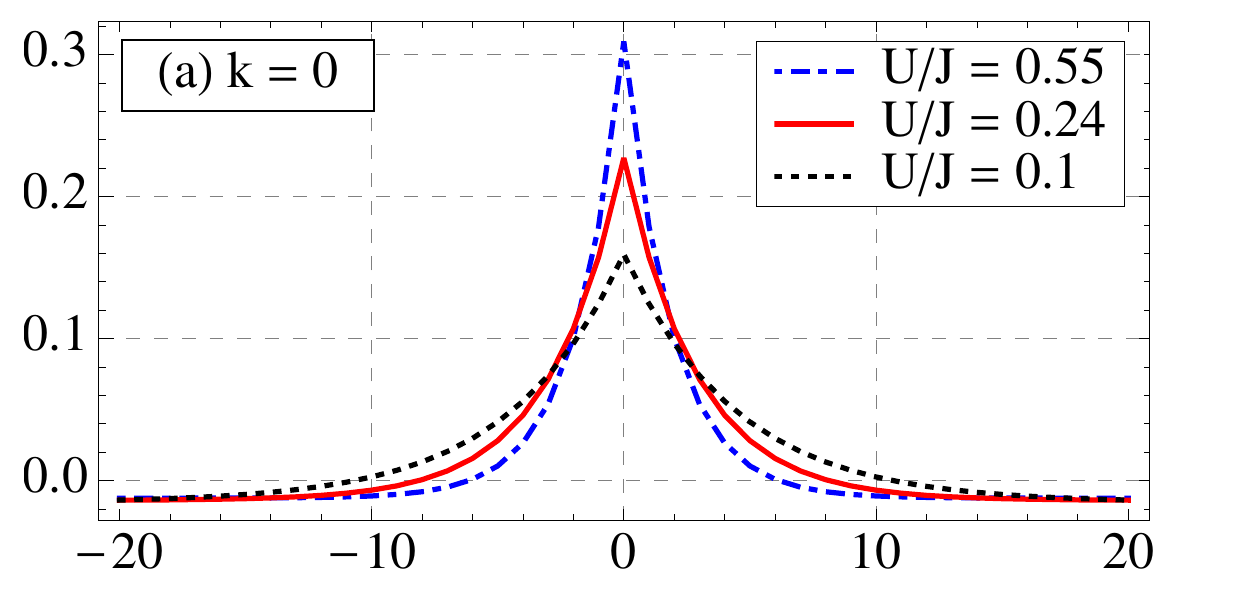}
  \put(-2.4,-36){\includegraphics[width=80.1\unitlength, height=38\unitlength]{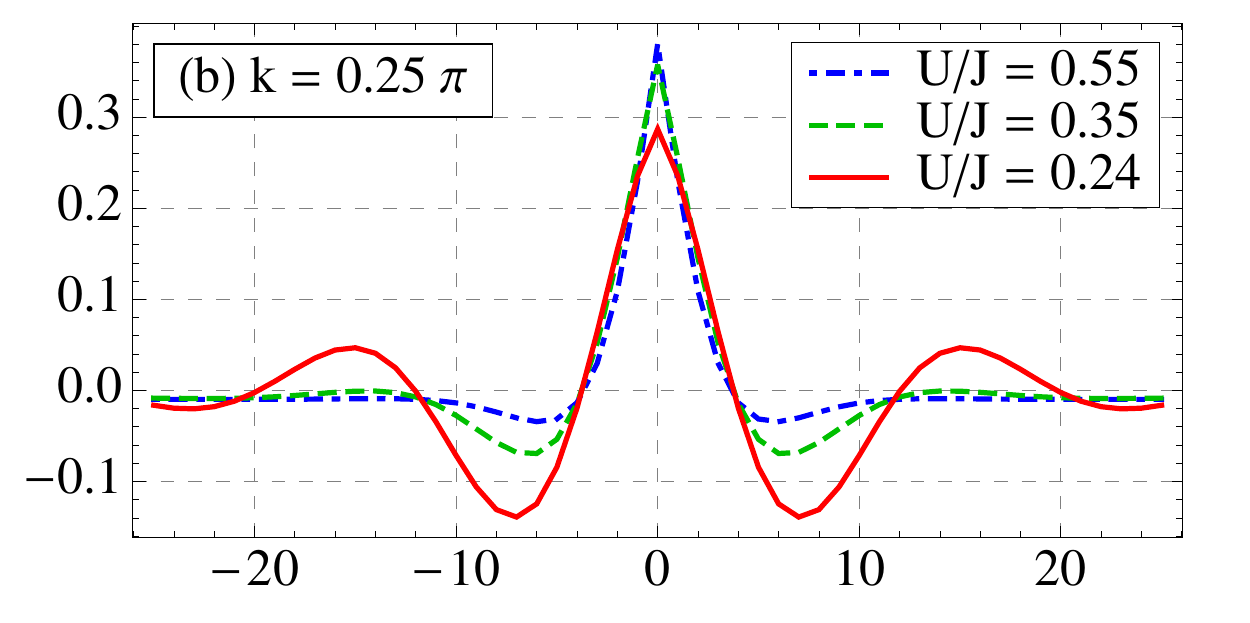}}
  \put(-2.3,-72.4){\includegraphics[width=78.6\unitlength, height=37.6\unitlength]{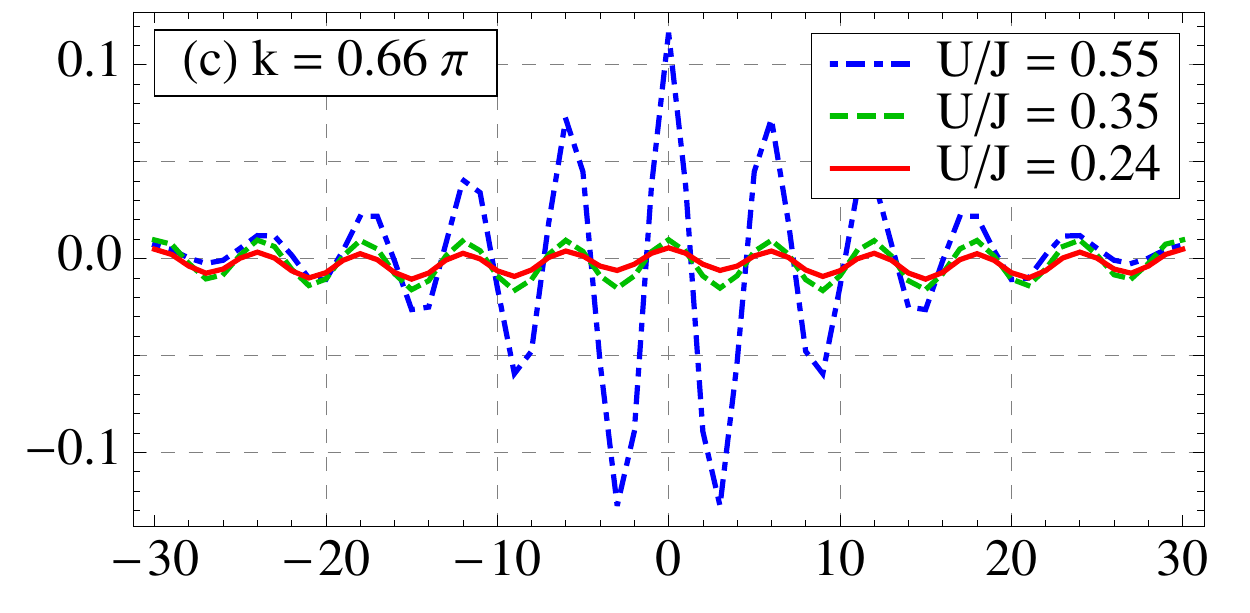}}
  \put(-2.1,-108.7){\includegraphics[width=76.7\unitlength, height=36.85\unitlength]{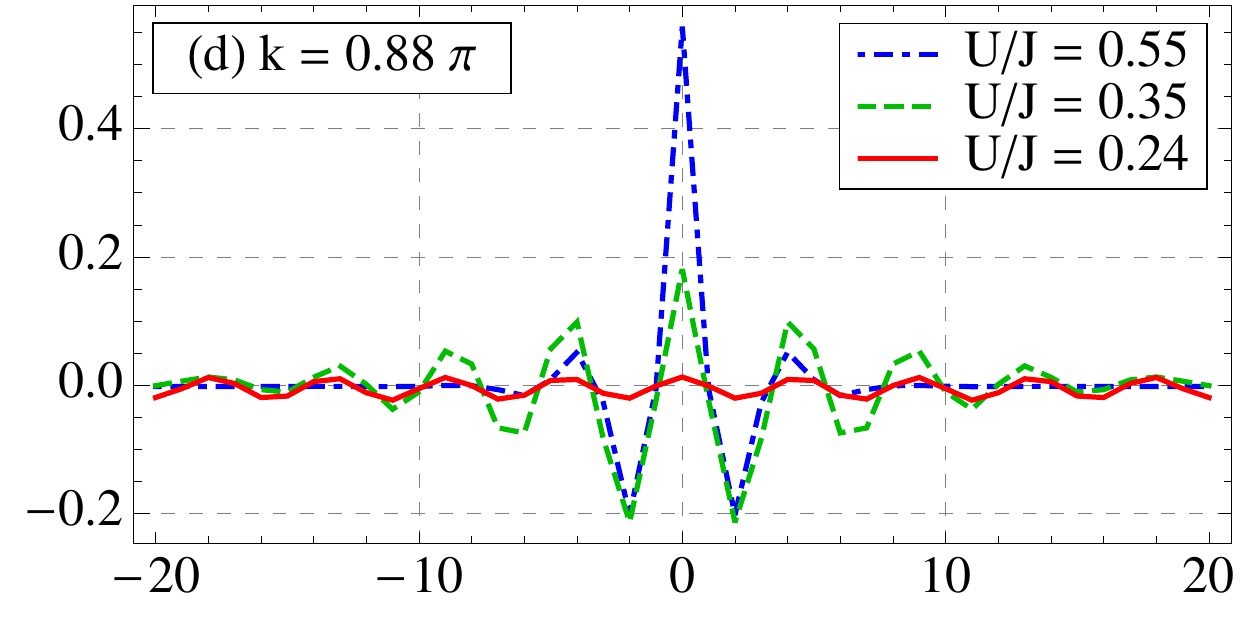}}
  \put(-4.5,-50.5){\rotatebox{90}{Correlation-hole density}}
  \put(35,-110.2){lattice sites}
\end{overpic}
\vspace{109.75\unitlength}
\caption{(Color online) Correlation-hole density in the vicinity of the impurity located at 0 for (a) $k=0$, (b) $k=0.25\pi$, (c) $k=0.66\pi$, and (d) $k=0.88\pi$, from our variational wavefunction. At strong interactions, we see polaronic excitation for all values of $k$. Here the impurity displaces nearby bath atoms, creating a (symmetric) bath density oscillation of period $\approx 4\pi/k$ within the healing length. The healing length increases with decreasing $U/J$, and is largest for $k \approx 2\pi/3$. For nonzero $k$, the system crosses over to the particle-hole continuum below a certain interaction strength, where the bath distribution becomes essentially independent of the impurity location. This crossover occurs at $U/J \approx$ 0.16, 0.44, and 0.29 for $k = 0.25\pi$, 0.66$\pi$, and 0.88$\pi$ respectively. Such a crossover does not happen for $k=0$ (compare with Fig. \ref{hole density bogoliubov}).}
\label{polaron plots}
\end{figure}

We find that the system exhibits stable polaronic excitations for all momenta at sufficiently strong repulsive interactions ($U/J \gtrsim 0.44$). Here the impurity displaces bath atoms around it, as illustrated by the correlations plotted in Fig. \ref{polaron plots}. The polaron becomes more spread out as $U/J$ is lowered. The momentum dependence of the polaron's size is more complicated. For a given $U/J$, the healing length increases with $k$ for small $k$, reaches a maximum for $k \approx 2\pi/3$, then decreases rapidly. At finite $k$ we observe decaying oscillations in the correlations with wavelength $\lambda \approx 4\pi/k$.

\begin{figure}[t]
\raggedleft
\begin{overpic}[trim=0cm 0cm 0.15cm 0cm, clip=true, width=82.6\unitlength, height= 38\unitlength ,tics=5]{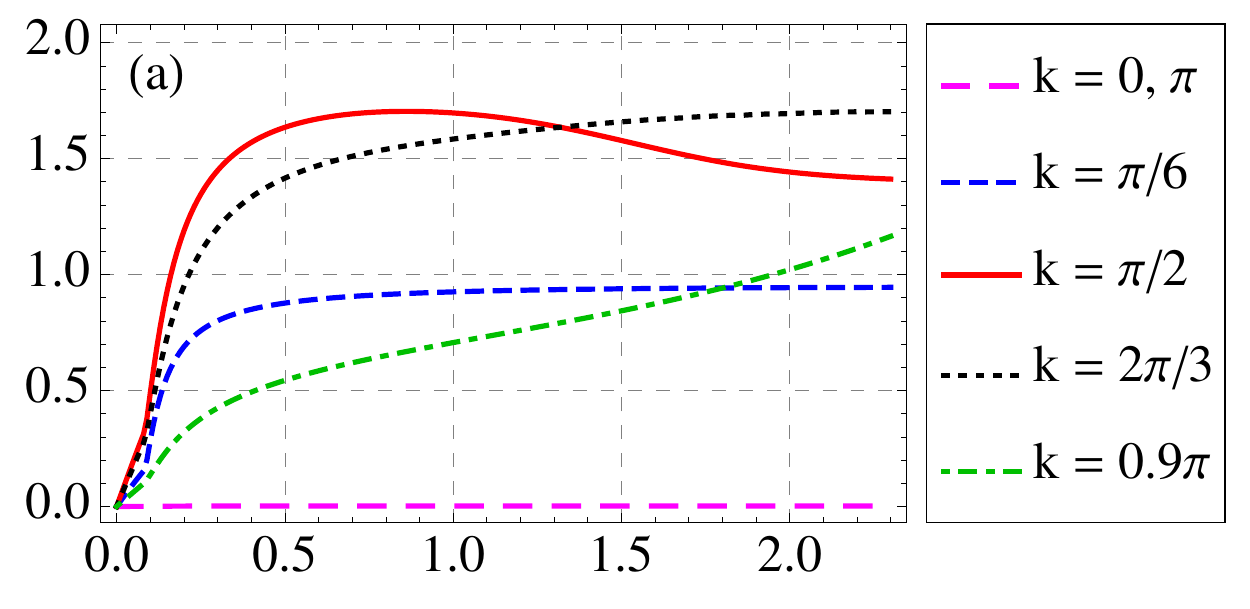}
  \put(30.5,-1.58){J/U}
  \put(1.73,-38.6){\includegraphics[width=80.52\unitlength, height=35.4\unitlength]{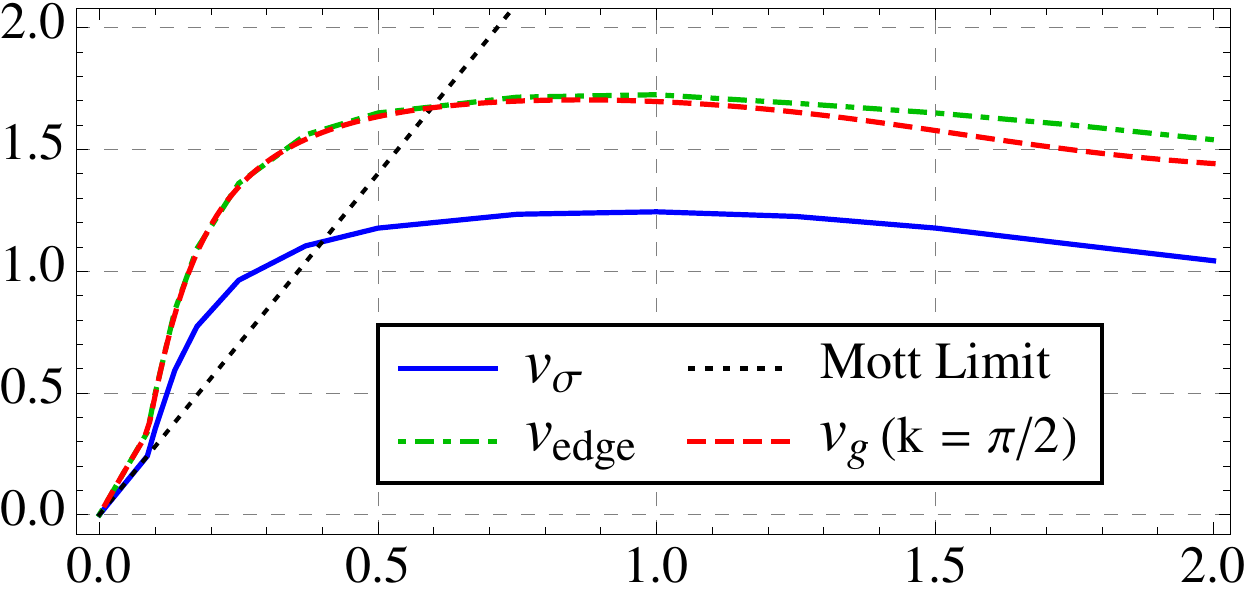}}
  \put(1.3,-78.225){\includegraphics[width=42\unitlength, height=34.75\unitlength]{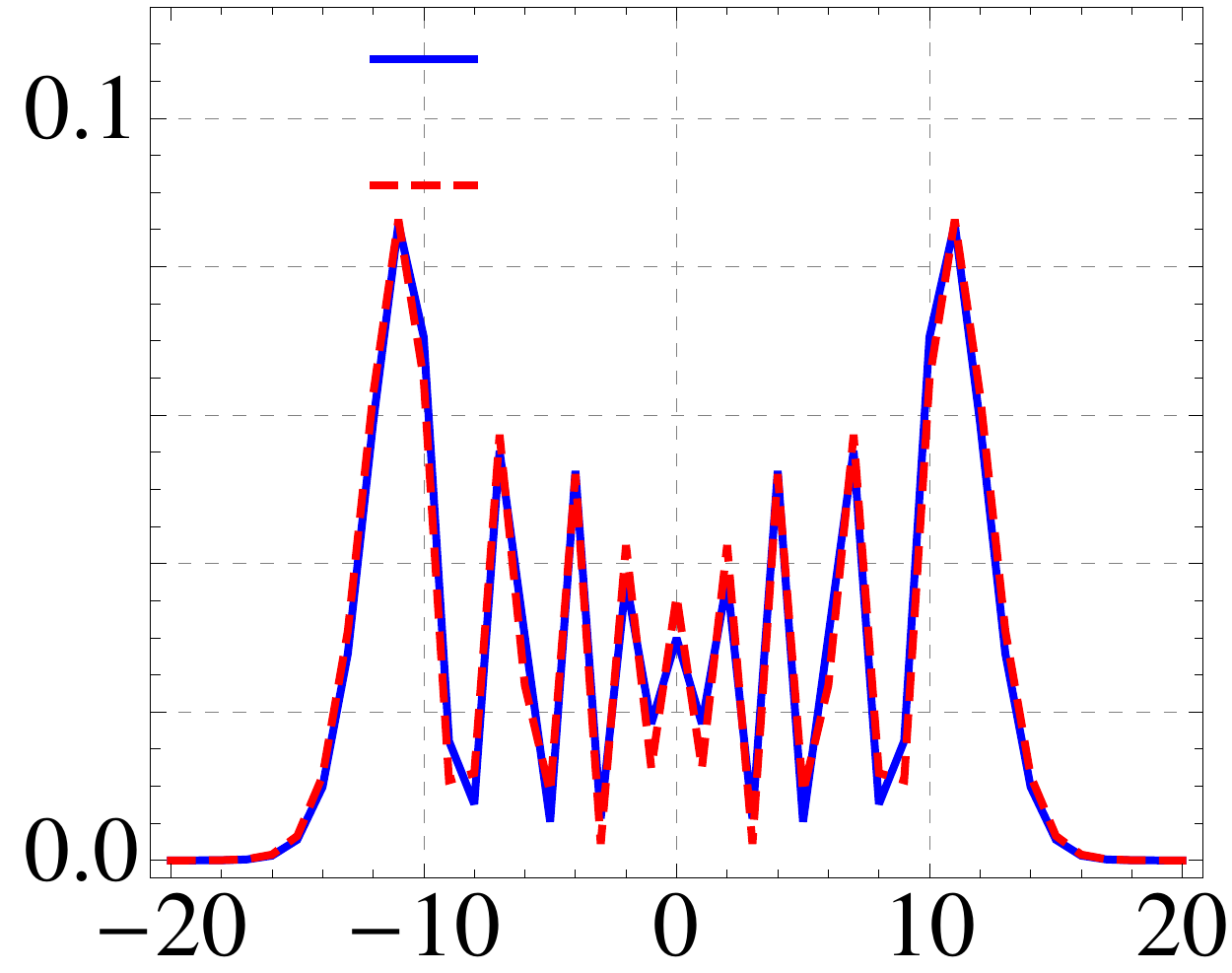}}
  \put(44.25,-77.65){\includegraphics[width=38.25\unitlength, height=34.15\unitlength]{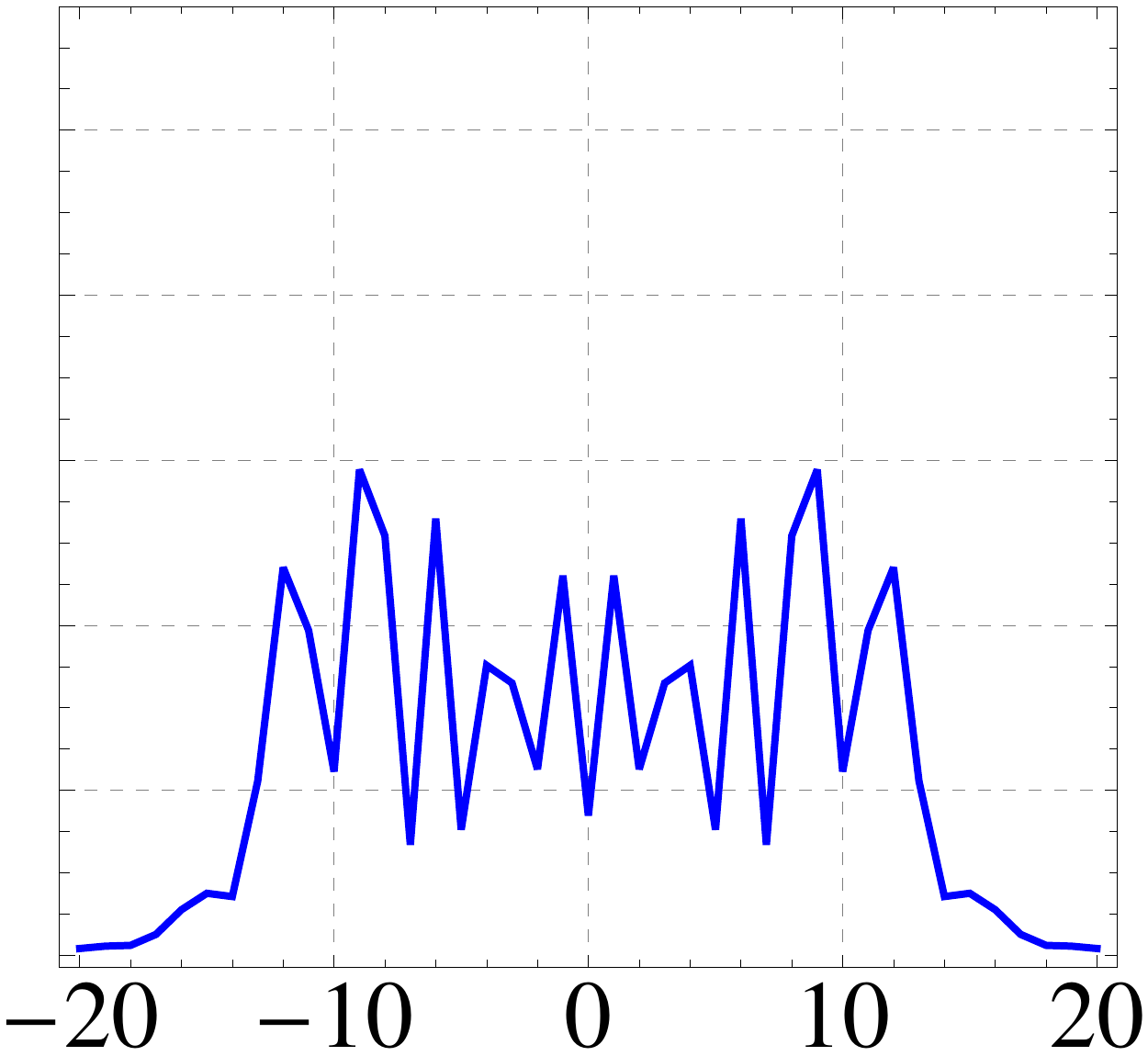}}
  \put(-3.3,11.7){\rotatebox{90}{$v_g(J/U,k)/J$}}
  \put(-3.23,-21.4){\rotatebox{90}{$v/J$}}
  \put(41,-41.1){J/U}
  \put(8.3,-8.3){(b)}
  \put(8,-46.75){(c) \hspace{0.40cm} $P_{\mathbf{0.05}}(j,t \hspace{-0.05cm}=\hspace{-0.05cm}64/J)$}
  \put(18.3,-50.75){$[\mathcal{J}_j (0.2 \hspace{-0.08cm}\times \hspace{-0.08cm}64)]^2$}
  \put(47.5,-46.75){(d)}
  \put(54.65,-51.45){$P_{\mathbf{2}}(j,t \hspace{-0.05cm}=\hspace{-0.05cm}8/J)$}
  \put(-3.1,-65.3){\rotatebox{90}{$P_{J/U}(j,t)$}}
  \put(34,-81.7){$j$ : lattice sites}
\end{overpic}
\vspace{81.1\unitlength}
\caption{(Color online) (a) Polaron group velocity at different momenta. After a rapid growth for small $J/U$, these saturate for $J/U \gtrsim 0.5$. (b) Propagation speed of an initially localized impurity in $\vert \psi (0) \rangle = \hat{b}_{0,\uparrow} \hat{b}_{0,\downarrow}^{\dagger} \vert MF \rangle$ projected into our variational subspace. $v_{\sigma}$ and $v_{\mbox{\scriptsize{edge}}}$ denote expansion speeds of the standard deviation and the leading edges of the impurity distribution respectively. $v_{\mbox{\scriptsize{edge}}}$ closely mimics the group velocity at $k=\pi/2$. For $J/U \gtrsim 0.5$, both speeds level off at values much smaller than the free-particle tunneling $2J$. (c), (d) Impurity distribution for $J/U = 0.05$ and $2$ respectively. In the Mott phase the distribution is described by a squared Bessel function, as predicted by the Heisenberg model. Whereas for large $J/U$ it has a distinctly different shape.}
\label{polaron speed}
\end{figure}
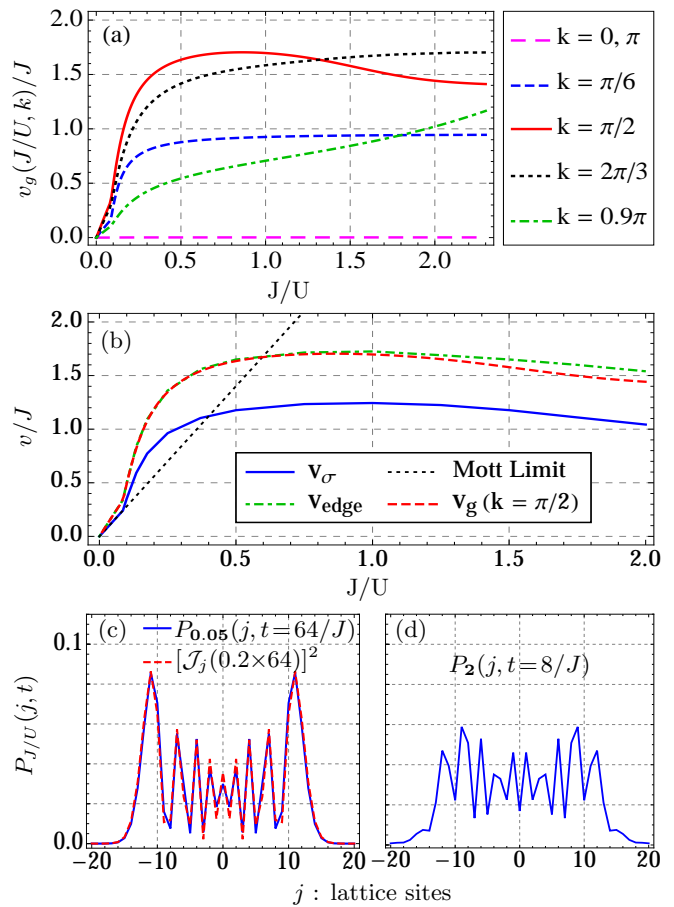

In Ref. \cite{fukuhara} the experimentalists measure the speed of propagation of an initially localized spin impurity. As a first step towards understanding such transport, in Fig. \ref{polaron speed}(a) we plot the polaron group velocity $v_g = \partial E_{\mbox{\scriptsize{var}}}(J/U,k) / \partial k$ for several points in the Brillouin zone. We see that the velocities (in units of $J$) rapidly grow for small $J/U$, then reach plateaus when $J \gtrsim 0.5 \;U$. The maximum velocity is much smaller than the maximum speed of propagation of a free particle with a tight-binding dispersion, $v_f = 2 J$.

To model the propagation of an initially localized impurity we project the initial state $\vert \psi (0) \rangle = \hat{b}_{0,\uparrow} \hat{b}_{0,\downarrow}^{\dagger} \vert MF \rangle$ into our variational subspace to find its time evolution:
\begin{equation}
\vert \psi (t) \rangle = \sum_k \frac{\langle k \vert \psi (0) \rangle}{\langle k \vert k \rangle} \vert k \rangle e^{-i E_{\mbox{\scriptsize{var}}}(k) t}\;.
\label{time evolution}
\end{equation}
The probability distribution of the impurity is calculated as $P(j,t)=\langle \psi (t) \vert \hat{b}_{j,\downarrow}^{\dagger} \hat{b}_{j,\downarrow} \vert \psi (t) \rangle / \langle \psi (t) \vert \psi (t) \rangle$. In Fig. \ref{polaron speed}(b) we plot the speed of propagation, $v_{\sigma}$, defined by taking the slope of $\sigma(t)$, where $\sigma(t)=\sum_j j^2 P(j,t)$. We find that for sufficiently large $t$, $\sigma$ increases linearly, and this speed is well-defined. In the Mott phase we find excellent agreement with the Heisenberg model which predicts $P(j,t)=[\mathcal{J}_j (J_{\mbox{\scriptsize{ex}}} t)]^2$, where $\mathcal{J}_j$ denotes the Bessel function of the first kind \cite{Mott probability} (see Fig. \ref{polaron speed}(c)). The distribution deviates more and more from this shape as $J/U$ increases (Fig. \ref{polaron speed}(d)). In addition to $v_{\sigma}$, we calculate the speed of propagation of the leading edge by fitting a Bessel function to the tail of the wave packet. We plot this speed in Fig. \ref{polaron speed}(b), finding that it closely follows the group velocity of the dispersion at $k=\pi/2$. This correspondence is consistent with the idea that the speed of the edge is constrained by the maximum group velocity (which is approximately the group velocity at $k=\pi/2$) \cite{cheneau}. Both $v_{\sigma}$ and $v_{\mbox{\scriptsize{edge}}}$ grow linearly with $J/U$ in the Mott regime, and become fairly flat well-inside the superfluid regime, in agreement with the experimental and simulation studies in Ref. \cite{fukuhara}. We find a kink at the phase transition point. We do not know if this kink is an artifact of the mean-field theory. No such feature is seen in the experiments. We find that the localized impurity state has less overlap with the variational subspace at larger $J/U$. This becomes especially important for $J/U \gtrsim 2.3$ when polarons become unstable for some momenta. Beyond this point the impurity dynamics are not well described by a single velocity.
\subsection{Crossover to the particle-hole continuum}\label{crossover section}
\begin{figure}[h]
\centering
\begin{overpic}[trim=0cm 0cm 0cm 0cm, clip=true, width=72\unitlength]{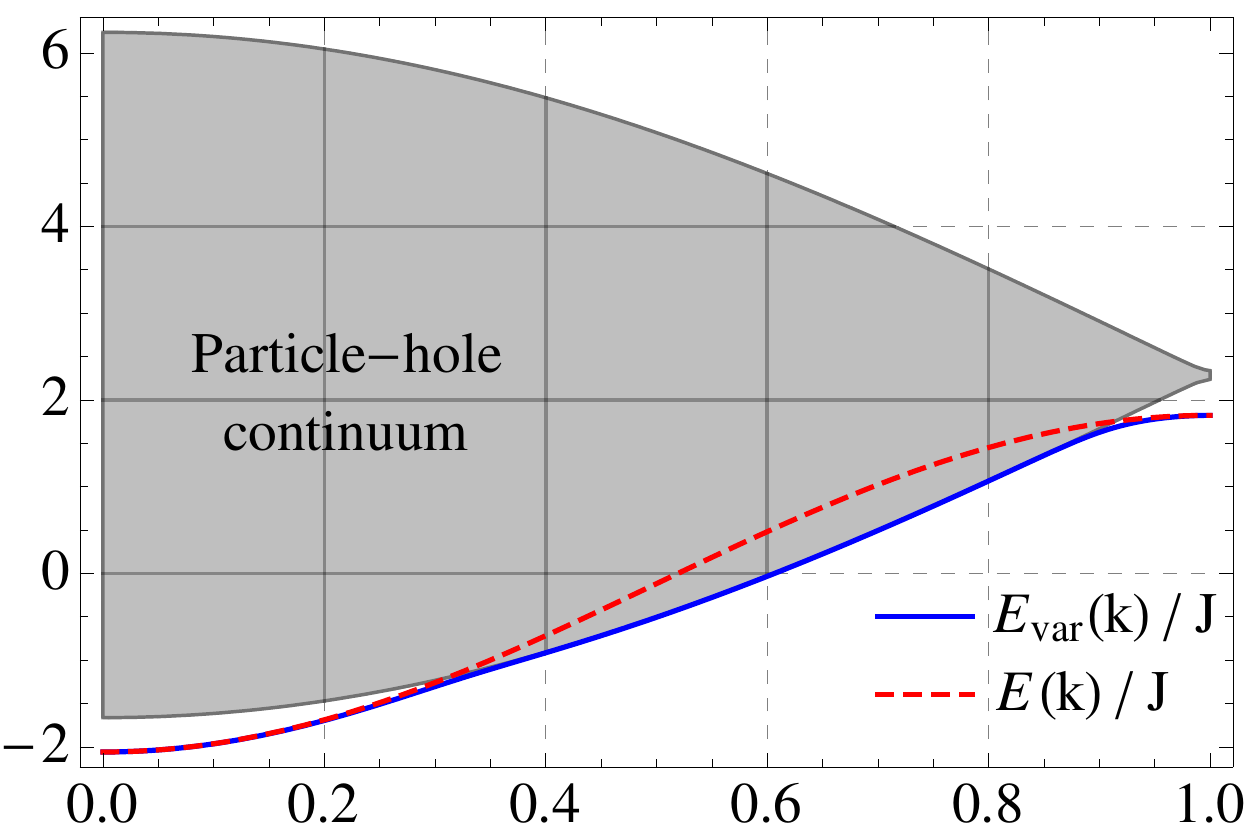}
  \put(-4.5,23.5){\rotatebox{90}{E/J}}
  \put(35.7,-3.8){k/$\pi$}
\end{overpic}
\vspace{1.5\unitlength}
\caption{(Color online) Energy of states with one impurity and one excess hole for $U/J = 0.37$. Solid line: variational ground state, $E_{\mbox{\scriptsize{var}}} (k)$. Gray region: independent particle-hole continuum. Dashed line: Approximate polaron dispersion $E (k) = E_0 - 2 J_{\mbox{\scriptsize{eff}}} \cos (k)$, where $E_0$ and $J_{\mbox{\scriptsize{eff}}}$ are chosen so that $E(0) = E_{\mbox{\scriptsize{var}}} (0)$ and $E (\pi) = E_{\mbox{\scriptsize{var}}} (\pi)$. At small\hspace{0.08cm}/\hspace{0.08cm}large $k$, $E_{\mbox{\scriptsize{var}}} (k)$ describes a stable polaron. For intermediate $k$, the polaron energy lies within the particle-hole continuum. Thus we expect it to be short-lived due to Landau damping.}
\label{continuum plot}
\end{figure}

As illustrated in Fig. \ref{polaron plots}, for weaker interactions the correlations between the impurity and the bath no longer decay. This indicates that the impurity and hole are not bound. To investigate this physics we study the wavefunction
\begin{equation}
\vert k\hspace{0.035cm};\hspace{0.02cm} p \rangle =  \hat{b}_{p,\uparrow} \hat{b}_{p-k,\downarrow}^{\dagger} \vert MF \rangle\;,
\label{two-particle wavefunction}
\end{equation}
where $p$ is a variational parameter. This represents an uncorrelated impurity and hole. It is a special case of Eq. (\ref{variational wavefunction}). For a given $k$, we have a continuum of energies $E_{\mbox{\scriptsize{two}}} (k,p)$ found by varying $p$. In Fig. \ref{continuum plot} we plot this continuum and our variational ground state energy for $U/J = 0.37$. For small and large values of $k$, the ground state energy is below the continuum, representing a stable polaron. At intermediate $k$, our variational approach finds the state at the bottom of the continuum, which does not correspond to a polaron. If the polaron exists at these momenta, its energy would be within the continuum. We expect that due to Landau damping it would have a short lifetime \cite{mahan}. We find that at small\hspace{0.08cm}/\hspace{0.08cm}large $k$, the polaron dispersion $E_{\mbox{\scriptsize{var}}} (k)$ is well approximated by the free particle form $E (k) = E_0 - 2 J_{\mbox{\scriptsize{eff}}} \cos (k)$, which Fig. \ref{continuum plot} shows entering the particle-hole continuum. 

\begin{figure}[h]
\includegraphics[width=0.98\columnwidth, height=0.56\columnwidth]{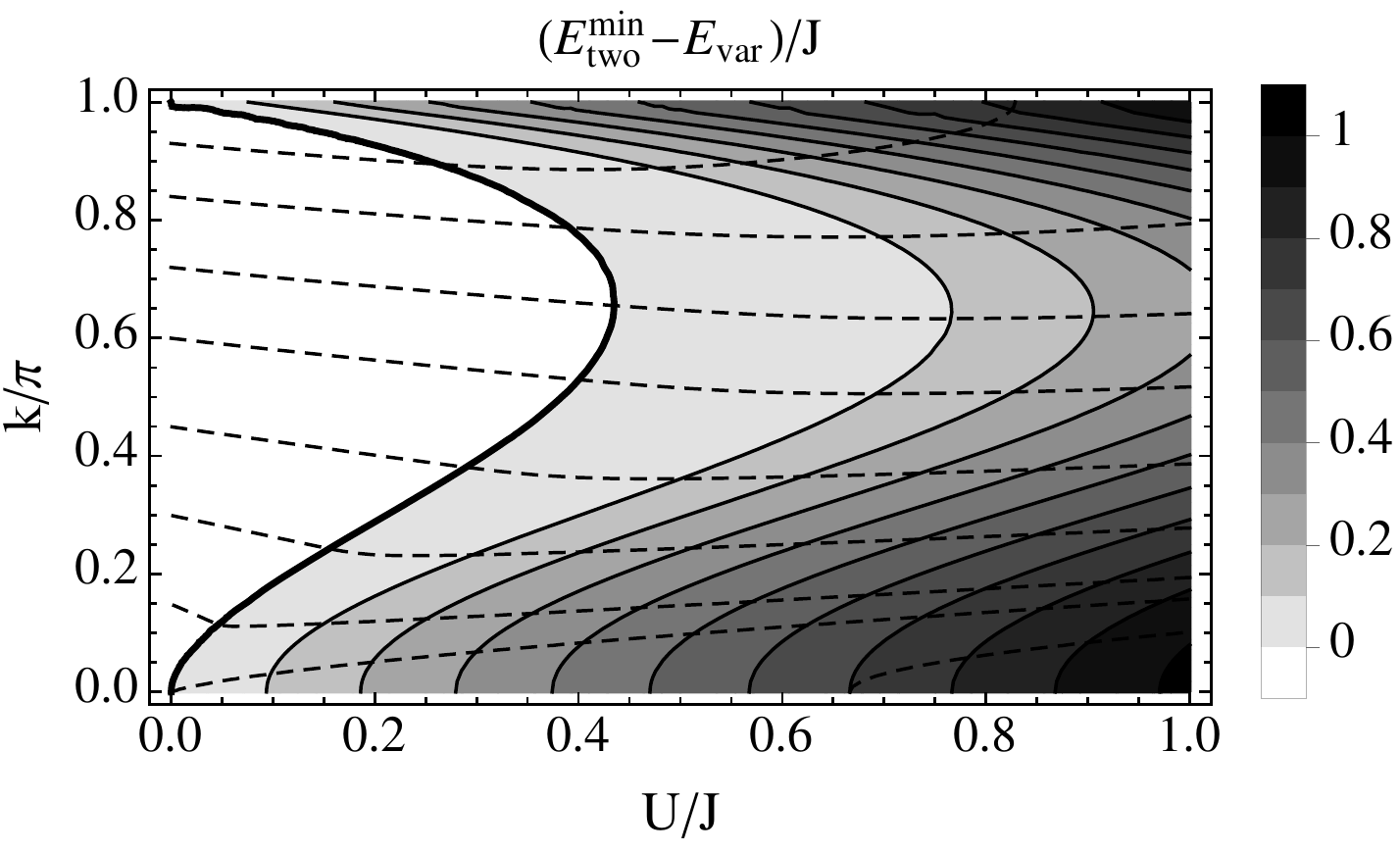}
\vspace{-0.3cm}
\caption{Contour plot of the energy difference between our variational state in Eq. (\ref{variational wavefunction}), and the bottom of the uncorrelated particle-hole continuum of states from Eq. (\ref{two-particle wavefunction}). Dotted lines show constant energy contours for $E_{\mbox{\scriptsize{var}}}$. For $U/J \gtrsim 0.44$, the variational ground state is lower in energy and describes a stable polaron. The two energies coincide to the left of the dark contour. Thus at weaker interactions there exist a growing range of momenta where the polaron is unstable, and the ground state belongs to the particle-hole continuum.}
\label{energy crossover}
\end{figure}

We denote the bottom of the particle-hole continuum as $E_{\mbox{\scriptsize{two}}}^{\mbox{\scriptsize{min}}} (J/U,k)$. In Fig. \ref{energy crossover} we estimate the region of the stability of the polaron by plotting the difference between the energies $E_{\mbox{\scriptsize{two}}}^{\mbox{\scriptsize{min}}}$ and $E_{\mbox{\scriptsize{var}}}$. The unstable region is to the left of the dark contour in Fig. \ref{energy crossover}, where these two energies are nearly equal. The instability window starts from $k \approx 2\pi/3$ at $U/J \approx 0.44$, and grows as the interaction is reduced.

\begin{figure}[h]
\includegraphics[width=0.98\columnwidth]{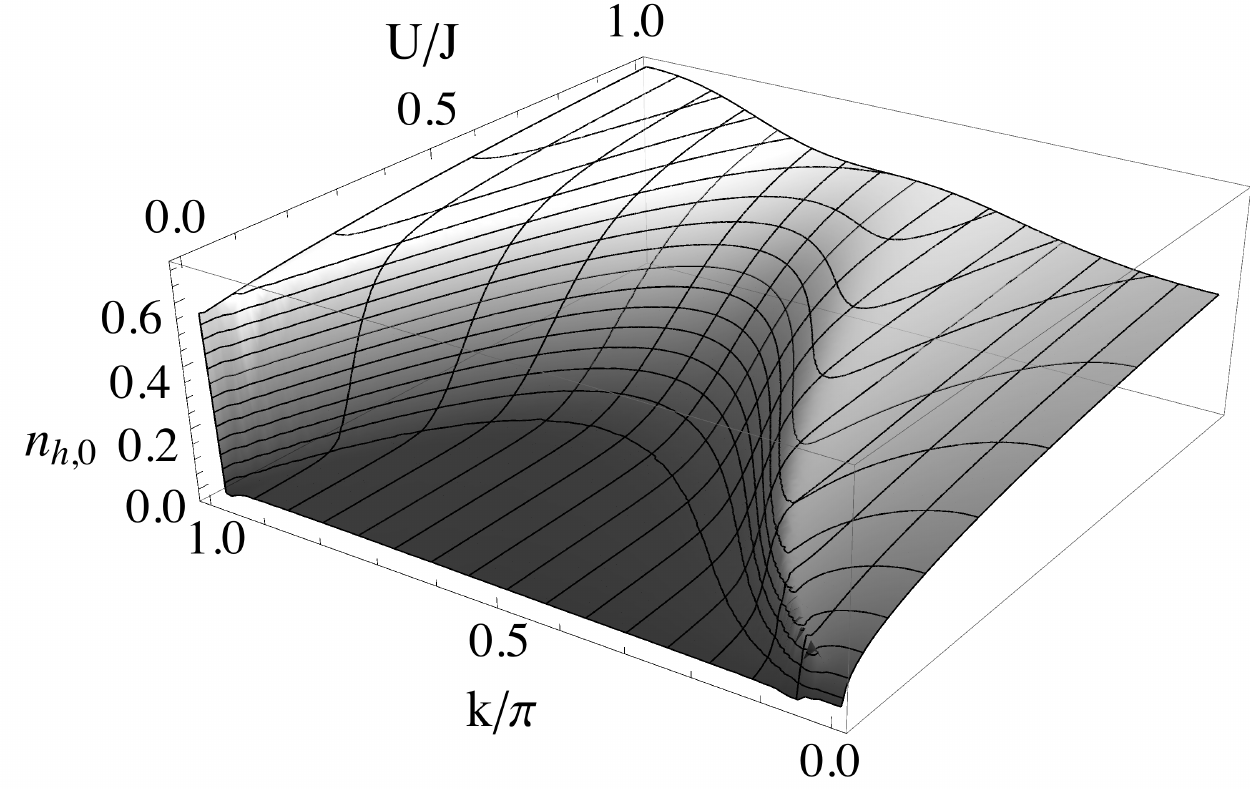}
\vspace{-0.3cm}
\caption{Correlation-hole density at the impurity site. As the system crosses over from the polaronic to the two-particle regime, the hole density rapidly falls toward zero. These correlations can be measured in experiments.}
\label{hole density crossover}
\end{figure}

To further illustrate this physics, in Fig \ref{hole density crossover} we plot $n_{h,0}$, the density of excess holes at the impurity site. We again see two distinct regions: the polaronic regime where $n_{h,0}$ is finite, and a two-particle regime where $n_{h,0}$ vanishes. The crossover location coincides with the dark curve in Fig. \ref{energy crossover}. These correlations could readily be measured in an experiment.

Throughout the two-particle regime, the lowest energy continuum state has $p \approx k/2$, leading to the small amplitude ($\sim 1/\mathcal{N}$) density oscillations of period $4\pi/k$ in Fig. \ref{polaron plots}. We can analytically calculate this optimal $p$ in the limit $U/J \to 0$. Here the Bogoliubov quasiparticle spectra reduce to the free-particle spectrum, $\varepsilon_0(p)=2 J (1-\cos p)$, and the quasiparticle operators are simply the particle and hole operators (Eqs. (\ref{epsilon0})$-$(\ref{uk and vk})). Since $\cos p + \cos (k-p)$ is maximized when $p=k/2$, it becomes energetically favorable to divide the total momentum equally between the impurity and hole.
\section{Two impurities and bipolarons}\label{bipolaron section}
A recent experimental study observed two-magnon bound states in the Mott regime \cite{fukuhara2}. Here the attraction arises from the fact that two flipped spins lower energy by sitting next to one another in the Heisenberg model. The stability of bipolarons in the superfluid phase is not obvious, and has not previously been explored in detail. Here we find that bipolarons are stable for $J/U \lesssim 0.15$, but unstable for weaker interactions.

\begin{figure}[h]
\raggedleft
\begin{overpic}[trim=0cm 0cm 0.5cm 0cm, clip=true, width=81.5\unitlength, height= 45\unitlength]{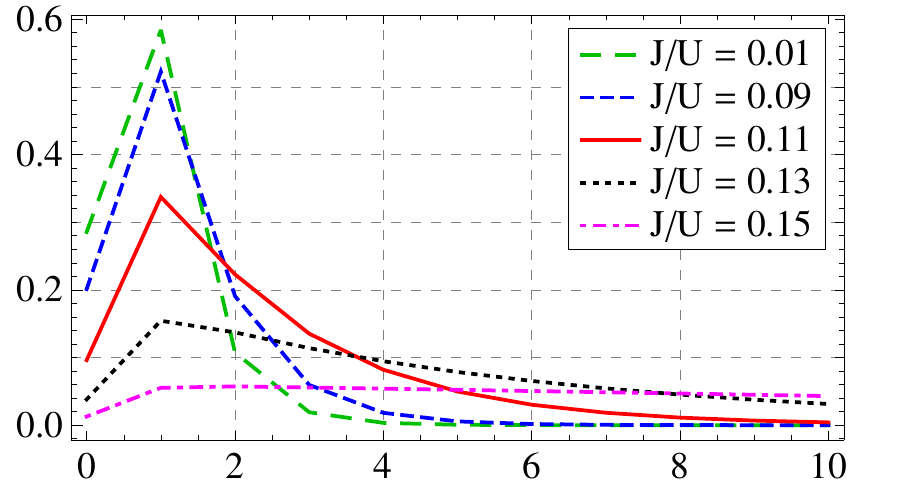}
  \put(-3.5,22.25){\rotatebox{90}{P({\it \hspace{-0.05cm}d})}}
  \put(42.75,-1.9){{\it d}}
\end{overpic}
\caption{(Color online) Separation probability of the two impurities, as predicted by the variational wavefunction in Eq. (\ref{bipolaron wavefunction}) on a lattice with 40 sites. In the Mott and the strongly interacting superfluid phase, the most probable separation of the impurities is one site, and the probabilities fall roughly exponentially with distance. For weaker interactions, the probabilities do not decay.}
\label{bipolaron plot}
\end{figure}

We study the following variational wavefunction for the case of zero total momentum, which is a simple extension of our model in Eq. (\ref{variational wavefunction}):
\begin{equation}
\vert \psi \rangle = \sum_{d \geq 0, \hspace{0.05cm}j} \Big[A(d) + \sum_{l} g(d,l) \;\hat{b}_{j+l,\uparrow}\Big]\hspace{0.05cm} \hat{b}_{j,\downarrow}^{\dagger} \hat{b}_{j+d,\downarrow}^{\dagger} \vert MF \rangle\;,
\label{bipolaron wavefunction}
\end{equation}
where $A(d)$ and $g(d,l)$ are variational parameters that control how the two impurities bind with holes and with each other. In Fig. \ref{bipolaron plot} we plot
\begin{equation}
P(d) = \sum_j \langle \psi \vert \hspace{0.03cm}\hat{b}_{j+d,\downarrow}^{\dagger} \hat{b}_{j,\downarrow}^{\dagger} \hat{b}_{j,\downarrow} \hat{b}_{j+d,\downarrow} \vert \psi \rangle / \langle \psi \vert \psi \rangle
\label{separation probability}
\end{equation}
for optimal parameter values, which gives the separation probability of the two impurities.

For sufficiently strong interactions, the probability peaks at unity separation, falling off rapidly for greater distances. This indicates that the two polarons are bound. As $J/U$ is raised, the distribution becomes flatter, so the average distance between the two polarons grows. For $J/U \gtrsim 0.15$, the average separation scales with the system size. We interpret this to mean that the polarons are no longer bound, and we are studying scattering states. Note that the Mott-superfluid transition occurs at $J/U \approx 0.086$ in our model, and our model gives stable polarons at all $k$ for $J/U \lesssim 2.3$. Thus we have four regions: (i) Mott (where polarons and bipolarons are stable), (ii) Superfluid with stable polarons and bipolarons, (iii) Superfluid with stable polarons but no bipolarons, and (iv) Superfluid where polarons are stable only for a narrow momentum range.
\vspace{-0.2cm}
\section{Summary and Outlook}\label{conclusions}
In this work we have studied spin impurities in a 1D Bose lattice gas through a computationally tractable variational ansatz. This ansatz provides an intuitive picture of phenomena seen in recent experiments and simulations. Our method reproduces the correct analytic results at strong and weak coupling.

For the case of a single impurity, we find stable polarons for all momenta when $U/J \gtrsim 0.44$. The polaron becomes larger with decreasing $U/J$. A moving polaron is bigger than a static one, attaining maximum size for $k \approx 2\pi/3$. We find that the impurity-hole correlations oscillate with wavelength $\approx 4\pi/k$. We calculate the impurity mobility from the polaronic dispersion. In the Mott phase, it increases linearly with $J/U$, as predicted by the Heisenberg model, whereas well-inside the superfluid phase, it saturates at a value much smaller than the free-particle hopping, as was experimentally observed in Ref. \cite{fukuhara}. At weaker interactions our model suggests that the polaron energy lies within the particle-hole continuum for intermediate $k$. Here we expect the polaron to be short-lived due to Landau damping. For the two-impurity system, we find stable bipolarons for $J/U \lesssim 0.15$.

Future experiments can probe the transition from the polaronic to the two-particle regime by studying impurity-hole correlations. As was illustrated in Ref. \cite{fukuhara}, one can measure the density at the impurity site, and compare it with the average density. This crossover should also show up in momentum resolved RF spectroscopy or other techniques which probe the single particle spectral function. The spectrum should be bimodal, with one peak coming from the polaron, and the other from the particle-hole continuum. This intuition is confirmed by explicit calculations in related systems \cite{steffen}. The techniques in Ref. \cite{fukuhara2} can be extended to study the stability of bipolarons in the superfluid phase. On the theoretical side, it would be interesting to study the system's behavior at higher dimensions and at filling factors different from unity \cite{ronzheimer}, as well as the effects of disorder on the polaron dynamics \cite{damski}. One of the most intriguing results we find is a kink in the polaron spread velocity when one crosses the Mott transition. It would be valuable to learn if this is an artifact or a real physical feature.
\vspace{-0.35cm}
\section*{Acknowledgements}\label{acknowledgements}
\vspace{-0.1cm} We thank Yariv Yanay for discussions. This material is based upon work supported by the National Science Foundation under Grant No. PHY-1068165. SD was partially supported by the Dr. V. Ramachandra Rao Summer Fellowship.
\appendix
\section{Eigenstates of the Hamiltonian in the Mott phase from second order perturbation theory} \label{appendix a}
For completeness, in this appendix we calculate the polaron states in the Mott limit to leading order in $J/U$ \cite{garcia-ripoll,giamarchi,kuklov,demler}. The Bose-Hubbard Hamiltonian in Eq. (\ref{bose-hubbard}) can be expressed as $\hat{H} = \hat{H}_0 - (J/U) \hat{H}_1$, where
\begin{align}
\hat{H}_0 &= \frac{U}{2} \sum_{l,\sigma,\sigma'} \hat{n}_{l,\sigma} \hat{n}_{l,\sigma'}\;,\label{H_0}\\
\hat{H}_1 &= U \hspace{-0.1cm} \sum_{(l_1,l_2),\sigma} \hat{b}_{l_1,\sigma}^{\dagger} \hat{b}_{l_2,\sigma}\;.\label{H_1}
\end{align}
We will treat $\hat{H}_1$ as a perturbation. A chemical potential is unnecessary as we will be working with states of fixed particle number. The eigenstates of the Heisenberg Hamiltonian (Eq. (\ref{heisenberg})) are given by $\vert k_{\mbox{\scriptsize{eff}}}\rangle = \sum_j e^{i k j} \;\vert\downarrow\rangle_j$, where $\vert\downarrow\rangle_j$ is the state where the impurity is localized at site $j$, and all other sites have one $\uparrow$ spin. We write the eigenstates of $\hat{H}$ as $\vert k \rangle = \vert k_{\mbox{\scriptsize{eff}}}\rangle + \sum_{\alpha} d_{\alpha} \vert \alpha \rangle$, where $\vert \alpha \rangle$ denotes states of the form
\begin{eqnarray}
\vert \beta \rangle_{ij} &=& \hat{b}_{j,\uparrow} \hat{b}_{i,\uparrow}^{\dagger} \vert\downarrow\rangle_i \;\;(i \neq j) \;,\label{betaij}\\
\vert \gamma \rangle_{ijk} &=& \hat{b}_{j,\uparrow} \hat{b}_{k,\uparrow}^{\dagger} \vert\downarrow\rangle_i \;\;(i\neq j \neq k) \;, \label{gammaijk}
\end{eqnarray}
which are parametrized by indices $ij$ and $ijk$. From degenerate second-order perturbation theory,
\begin{align}
\nonumber \vert k \rangle = &\;\vert k_{\mbox{\scriptsize{eff}}}\rangle + \frac{J}{U} \sum_j e^{i k j} \sum_{\alpha} \vert \alpha \rangle \frac{\langle \alpha \vert \hat{H}_1 \vert \downarrow \rangle_j}{\langle \alpha \vert \hat{H}_0 \vert \alpha \rangle - {}_j\langle \downarrow \vert \hat{H}_0 \vert \downarrow \rangle_j}\\
\nonumber = &\;\vert k_{\mbox{\scriptsize{eff}}}\rangle + \frac{J}{U^2} \sum_j e^{i k j} \sum_{\alpha} \vert \alpha \rangle \langle \alpha \vert \hat{H}_1 \vert \downarrow \rangle_j\\
\nonumber = &\hspace{0.05cm}\sum_j e^{i k j} \Big[\vert\downarrow\rangle_j + \frac{J}{U} \big\{(1 + e^{i k}) \vert + \rangle_j + (1 + e^{-i k}) \vert - \rangle_j \big\}\Big]\\
&\hspace{0.05cm}+\sqrt{2} \;\frac{J}{U} \sum_j e^{i k j} \hspace{-0.1cm}\sum_{l \neq j,j-1}\hspace{-0.1cm} \big(\vert \gamma \rangle_{j(l+1)l} + \vert \gamma \rangle_{jl(l+1)} \big)\hspace{-1cm}
\end{align}
where $\vert \pm \rangle_j = \hat{b}_{j\pm1,\uparrow} \hat{b}_{j,\uparrow}^{\dagger} \vert\downarrow\rangle_j$. The $k$ dependence in the dispersion comes from the matrix element of the Hamiltonian between $\vert k_{\mbox{\scriptsize{eff}}}\rangle$ and the states $\vert \pm \rangle_j$ which represent impurity hopping. The other correction states only contribute a constant term.
\section{Bogoliubov weak-coupling analysis} \label{appendix b}
In this appendix we calculate the correlation-hole density around an impurity within the Bogoliubov approximation. Using $\hat{b}_{0,\sigma} = \hat{b}_{0,\sigma}^{\dagger} = \sqrt{N^{\sigma}}$ in the Bose-Hubbard Hamiltonian (Eq. (\ref{momentum space hamiltonian})) and retaining quadratic fluctuations, we obtain the mean-field Hamiltonian:\\
\begin{align}
\nonumber \hat{H} = \;&H_0 - \sum_{p \neq 0, \sigma} \Big(2 J \cos p + \tilde{\mu} - U \frac{N}{\mathcal{N}}\Big) \;\hat{b}_{p,\sigma}^{\dagger} \hat{b}_{p,\sigma} \\
&+ \frac{U}{2} \hspace{-0.1cm}\sum_{p \neq 0, \sigma_1, \sigma_2}\hspace{-0.1cm} \Big[\sqrt{n^{\sigma_1} n^{\sigma_2}}\; \hat{b}_{p,\sigma_1}^{\dagger} (\hat{b}_{p,\sigma_2} + \hat{b}_{-p,\sigma_2}^{\dagger}) + \mbox{h.c.} \Big],
\label{mean field hamiltonian}
\end{align}
where $n^{\sigma} \hspace{-0.1cm}= \hspace{-0.1cm} N^{\sigma}/\mathcal{N}$, $N \hspace{-0.1cm}=\hspace{-0.1cm} \sum_{\sigma} N^{\sigma}$, $\tilde{\mu}\hspace{-0.05cm} =\hspace{-0.05cm} \mu - U/2$, and $H_0 = -(2 J + \tilde{\mu}) N \hspace{-0.03cm}+\hspace{-0.03cm} \frac{U}{2 \mathcal{N}} N^2$. The constant $H_0$ is minimized when
\begin{equation}
\tilde{\mu} = -2 J + U (N / \mathcal{N})\;.
\label{bogoliubov chemical potential}
\end{equation}
Substituting this back into Eq. (\ref{mean field hamiltonian}) yields
\begin{align}
\nonumber \hat{H} = &-\frac{1}{2} \frac{U}{\mathcal{N}} N^2 + \sum_{p \neq 0, \sigma} \varepsilon_0 (p) \;\hat{b}_{p,\sigma}^{\dagger} \hat{b}_{p,\sigma} \\
& + \frac{U}{2} \hspace{-0.1cm}\sum_{p \neq 0, \sigma_1, \sigma_2}\hspace{-0.1cm} \Big[\sqrt{n^{\sigma_1} n^{\sigma_2}}\; \hat{b}_{p,\sigma_1}^{\dagger} (\hat{b}_{p,\sigma_2} + \hat{b}_{-p,\sigma_2}^{\dagger}) + \mbox{h.c.} \Big],
\label{to be diagonalized}
\end{align}
where $\varepsilon_0 (p) = 2 J (1-\cos p)$. We wish to diagonalize this Hamiltonian to produce
\begin{equation}
\hat{H} = E_0 + \sum_{p \neq 0} \big(\varepsilon_c (p) \;\hat{c}_{p}^{\dagger} \hat{c}_p + \varepsilon_d (p) \;\hat{d}_{p}^{\dagger} \hat{d}_p \big)\;,
\label{diagonalized}
\end{equation}
where the quasi-particle operators $\hat{c}_p$ and $\hat{d}_p$ are related to $\hat{b}_{p,\sigma}$ by a Bogoliubov transformation. A convenient way to find this transformation is to analyze the Heisenberg equations of motion:
\begin{flalign}
&i \partial_t \hat{b}_{p,\sigma} = \varepsilon_0 (p) \hspace{0.05cm} \hat{b}_{p,\sigma} \hspace{-0.05cm}+ U \sqrt{n^{\sigma}} \sum_{\sigma'} \hspace{-0.1cm}\sqrt{n^{\sigma'}} (\hat{b}_{p,\sigma'} + \hat{b}_{-p,\sigma'}^{\dagger}) ,\label{eom bpsigma} \hspace{-0.5cm}&\\[-2.7mm]
&\hspace{0.25cm} i \partial_t \hat{c}_{p} = \varepsilon_c (p) \hspace{0.05cm} \hat{c}_{p}\;, \label{eom cp} &\\
&\hspace{0.22cm} i \partial_t \hat{d}_{p} = \varepsilon_d (p) \hspace{0.05cm} \hat{d}_{p}\;. \label{eom dp} &
\end{flalign}
These can be written more succinctly as
\begin{align}
i \partial_t \hat{B}_{p,\sigma}^+ &= \varepsilon_0 (p) \hat{B}_{p,\sigma}^- \;, \label{delt Bpsigmaplus}\\
i \partial_t \hat{B}_{p,\sigma}^- &= \varepsilon_0 (p) \hat{B}_{p,\sigma}^+ + 2 U \sqrt{n^{\sigma}} \sum_{\sigma'} \sqrt{n^{\sigma'}} \hat{B}_{p,\sigma'}^+ \;, \label{delt Bpsigmaminus}\\[-2.8mm]
i \partial_t \hat{C}_p^{\pm} &= \varepsilon_c (p) \hat{C}_p^{\mp} \;, \label{delt Cp}\\
i \partial_t \hat{D}_p^{\pm} &= \varepsilon_d (p) \hat{D}_p^{\mp} \;. \label{delt Dp}
\end{align}
where $\hat{B}_{p,\sigma}^{\pm} = \frac{1}{\sqrt{2}} (\hat{b}_{p,\sigma} \pm \hat{b}_{-p,\sigma}^{\dagger})$, $\hat{C}_{p}^{\pm} = \frac{1}{\sqrt{2}} (\hat{c}_{p} \pm \hat{c}_{-p}^{\dagger})$, and $\hat{D}_{p}^{\pm} = \frac{1}{\sqrt{2}} (\hat{d}_{p} \pm \hat{d}_{-p}^{\dagger})$. We define the transformation
\begin{equation}
\hat{B}_{p,\sigma}^{\pm} = \Gamma_{p,\sigma}^{\pm} \hat{C}_p^{\pm} + \Delta_{p,\sigma}^{\pm} \hat{D}_p^{\pm}\;.
\label{bogoliubov transformation appendix}
\end{equation}
From bosonic commutation relations it follows that
\begin{equation}
\Gamma_{p,\sigma}^+ \Gamma_{p,\sigma}^- + \Delta_{p,\sigma}^+ \Delta_{p,\sigma}^- = 0\;.
\label{commutation}
\end{equation}
In addition, using Eqs. (\ref{delt Cp})$-$(\ref{bogoliubov transformation appendix}) in Eqs. (\ref{delt Bpsigmaplus}) and (\ref{delt Bpsigmaminus}) yields
\begin{align}
\Gamma_{p,\sigma}^+ \;\varepsilon_c (p) &= \Gamma_{p,\sigma}^- \;\varepsilon_0 (p) \;, \label{Gammaplus equation}\\
\Gamma_{p,\sigma}^- \;\varepsilon_c (p) &= \Gamma_{p,\sigma}^+ \;\varepsilon_0 (p) + 2 U \sqrt{n^{\sigma}} \sum_{\sigma'}\hspace{-0.05cm} \sqrt{n^{\sigma'}} \hspace{0.05cm} \Gamma_{p,\sigma'}^+\;,
\label{Gammaminus equation}
\end{align}
and similar equations for $\Delta$. These equations, along with Eq. (\ref{commutation}), can be solved to obtain
\begin{align}
\varepsilon_c (p) &= \sqrt{(\varepsilon_0 (p))^2 + 2 \;\varepsilon_0 (p) \;U\; (n^{\uparrow} + n^{\downarrow})} \;, \label{epsilonc appendix}\\
\varepsilon_d (p) &= \varepsilon_0 (p) \;, \label{epsilond appendix}\\[-0.6mm]
\Gamma_{p,\uparrow/\downarrow}^+ &= \sqrt{f^{\uparrow/\downarrow}\; \varepsilon_0 (p)/\varepsilon_c (p)} \;, \label{Gammaplus solution}\\[-1mm]
\Gamma_{p,\uparrow/\downarrow}^- &= \sqrt{f^{\uparrow/\downarrow}\; \varepsilon_c (p)/\varepsilon_0 (p)} \;, \label{Gammaminus solution}\\[-1mm]
\Delta_{p,\uparrow/\downarrow}^{\pm} &= \mp \sqrt{f^{\downarrow/\uparrow}} \;, \label{Delta solution}
\end{align}
\noindent where $f^{\sigma} = N^{\sigma} / N$.\\

\vspace{-2.5mm} In the limit $n^{\downarrow} \to 0$ and $n^{\uparrow} \to 1$, we can calculate the correlation-hole density as (Eq. (\ref{cj})):
\begin{align}
\nonumber n_{h,j} &= \big(\langle \hat{n}_{j,\uparrow} \rangle \langle \hat{n}_{0,\downarrow} \rangle - \langle \hat{n}_{j,\uparrow} \hat{n}_{0,\downarrow} \rangle \big) / n^{\downarrow} \\
\nonumber &= \frac{1}{\mathcal{N}^2 \hspace{0.05cm} n^{\downarrow}} \sum_{p,q,s,t} \Big[ \langle \hat{b}_{p,\uparrow}^{\dagger} \hat{b}_{q,\uparrow} \rangle \langle \hat{b}_{s,\downarrow}^{\dagger} \hat{b}_{t,\downarrow} \rangle \\[-3.5mm]
& \hspace{2.5cm} - \langle \hat{b}_{p,\uparrow}^{\dagger} \hat{b}_{q,\uparrow} \hat{b}_{s,\downarrow}^{\dagger} \hat{b}_{t,\downarrow} \rangle \Big] e^{i(p-q)j}\;. \label{nhj appendix 1}
\end{align}
\noindent Replacing the zero-momenta operators by $\sqrt{N^{\sigma}}$ and keeping the quadratic terms,
\begin{align}
\nonumber n_{h,j} &= -\frac{\sqrt{N^{\uparrow} N^{\downarrow}}}{\mathcal{N}^2 \hspace{0.05cm} n^{\downarrow}} \sum_{p,q \neq 0} \langle (\hat{b}_{p,\uparrow} + \hat{b}_{-p,\uparrow}^{\dagger}) (\hat{b}_{q,\downarrow} + \hat{b}_{-q,\downarrow}^{\dagger})\rangle \;e^{-i p j} \\[-1mm]
&= -\frac{2 \sqrt{N^{\uparrow} N^{\downarrow}}}{\mathcal{N}^2 \hspace{0.05cm} n^{\downarrow}} \sum_{p,q \neq 0} \langle \hat{B}_{p,\uparrow}^+ \hat{B}_{q,\downarrow}^+ \rangle \;e^{-i p j}\;. \label{nhj appendix 2}
\end{align}
Substituting Eq. (\ref{bogoliubov transformation appendix}) in the above equation and using the fact that $\hat{c}_p \vert MF \rangle = \hat{d}_p \vert MF \rangle = 0$, we get
\begin{align}
\nonumber n_{h,j} &= -\frac{\sqrt{N^{\uparrow} N^{\downarrow}}}{\mathcal{N}^2 \hspace{0.05cm} n^{\downarrow}} \sum_{p \neq 0} \big(\Gamma_{p,\uparrow}^+ \Gamma_{p,\downarrow}^+ + \Delta_{p,\uparrow}^+ \Delta_{p,\downarrow}^+ \big) \\[-0.9mm]
&= \big(1/\mathcal{N}\big) \sum_{p \neq 0} \big(1 - \varepsilon_0 (p) / \varepsilon_c (p)\big) \cos pj\;. \label{nhj appendix 3}
\end{align}

\end{document}